\documentclass[sigconf]{acmart}
\usepackage{algorithm2e}
\usepackage{sidecap}
\sidecaptionvpos{figure}{c}
\RestyleAlgo{ruled}
\SetKwComment{Comment}{/* }{ */}
\LinesNumbered

\AtBeginDocument{%
  }

\copyrightyear{2023}
\acmYear{2023}
\setcopyright{rightsretained}
\acmConference[WWW '23 Companion]{Companion Proceedings of the ACM Web Conference 2023}{April 30-May 4, 2023}{Austin, TX, USA}
\acmBooktitle{Companion Proceedings of the ACM Web Conference 2023 (WWW '23 Companion), April 30-May 4, 2023, Austin, TX, USA}
\acmDOI{10.1145/3543873.3587570}
\acmISBN{978-1-4503-9419-2/23/04}

\begin{document}

\title{Graph Embedding for Mapping Interdisciplinary Research Networks}

\author{Eoghan Cunningham}
\orcid{0000-0002-0435-1962}
\affiliation{%
  \institution{School of Computer Science, University College Dublin}
  \country{Dublin, Ireland}
}
\email{eoghan.cunningham@ucdconnect.ie}
\author{Derek Greene}
\orcid{0000-0001-8065-5418}
\affiliation{%
  \institution{School of Computer Science, University College Dublin}
  \country{Dublin, Ireland}
}
\email{derek.greene@ucd.ie}

\renewcommand{\shortauthors}{Eoghan Cunningham \& Derek Greene}
\begin{abstract}
Representation learning is the first step in automating tasks such as research paper recommendation, classification, and retrieval. Due to the accelerating rate of research publication, together with the recognised benefits of interdisciplinary research, systems that facilitate researchers in discovering and understanding relevant works from beyond their immediate school of knowledge are vital. This work explores different methods of research paper representation (or document embedding), to identify those methods that are capable of preserving the interdisciplinary implications of research papers in their embeddings. In addition to evaluating state of the art methods of document embedding in a interdisciplinary citation prediction task, we propose a novel Graph Neural Network architecture designed to preserve the key interdisciplinary implications of research articles in citation network node embeddings. Our proposed method outperforms other GNN-based methods in interdisciplinary citation prediction, without compromising overall citation prediction performance.
\end{abstract}



\maketitle

\section{Introduction}
Globally, the rate of scientific publication is accelerating rapidly. Such unprecedented growth necessitates the development of tools to support the search and recommendation of scientific research papers. Many of these methods hinge on research paper representation learning (or `document embedding'). This refers to the task of discovering useful representations of scientific articles, typically for use in downstream applications like classification, prediction, and retrieval \cite{cohan2020specter}. These methods rely primarily on Natural Language Processing (NLP) and Network Analysis techniques to learn document embeddings according to a paper's content, a paper's citation relations, or some combination of these sources \cite{kozlowski2021semantic}.

In this work, we explore various approaches to scientific document representation from the perspective of mapping \textit{interdisciplinary} research. We posit that document embeddings which preserve a work's interdisciplinary implications (or applications) are more desirable for downstream tasks, such as article recommendation or retrieval. The benefits of interdisciplinary research are well established \cite{ledford2015team}. However, due to the sheer volume of research published each day, combined with the recognised cognitive strain associated with reading outside of ones chosen and practiced discipline \cite{wear1999challenges}, engaging broadly with research from multiple disciplines is challenging. Accordingly, systems for research paper recommendation or retrieval that allow researchers to discover relevant works from beyond their own disciplines or communities must be developed. To this end, we assess a number of document embedding methods with respect to their ability to recognise and preserve the interdisciplinary implications of a research paper in their learned representations. Further, we propose a novel Message-Passing Graph Neural Network architecture, designed specifically for embedding interdisciplinary research. Motivated by work which highlights a relationship between certain citation graph substructures and interdisciplinary research \cite{cunningham2022assessing, cunningham2022structure}, our method permits the aggregation of information from neighbouring nodes in a manner that is sensitive to the local graph structure and additional relations between the neighbours. We evaluate this new method, referred to as Community-Based Sample and Aggregation method (\textit{ComBSAGE}), via an interdisciplinary citation prediction task. The underlying hypothesis here is that those embeddings that are capable of recognising and preserving the \textit{interdisciplinarity} of research articles will better predict citations between two papers that are distant in either the semantic or relational space of research papers.   

\section{Background}

\begin{figure*}[h]
  \includegraphics[width=0.75\textwidth]{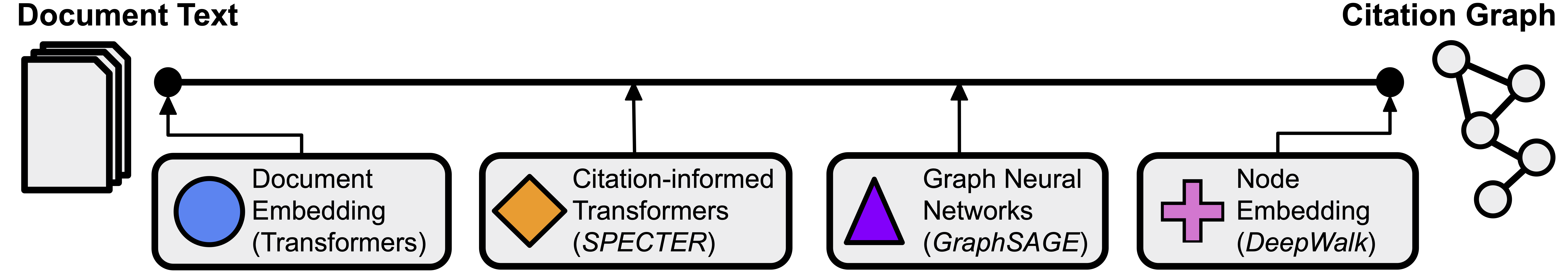}
  \caption{Spectrum of methods for research paper representation learning, with example methods listed.}
  \label{fig:spectrum}
\vskip-1em
\end{figure*}

Representation learning for scientific documents is the task of representing research papers in some dense vector space, such that the important similarities and relations between the papers are preserved i.e., 
semantically related papers should have similar representations (or \textit{embeddings}) \cite{kozlowski2021semantic}. Scientific articles may be related if they pertain to the same topic (i.e., have similar content), or if there is an application or transfer of knowledge from one to the other (i.e., a citation). Accordingly, article content and citation relations represent the primary sources of information for scientific document representation. We limit our analysis to methods that rely on these sources, and propose that all of the approaches we include can be considered on the spectrum illustrated in Figure \ref{fig:spectrum}. We place those methods that rely solely on article text on the left, those methods that rely only on the citation network on the right, and any combined or hybrid methods lie on the continuum in between.
In the rest of this section, we discuss some of the important approaches to scientific document representation according to their position on this spectrum. 

Large language models like \textit{BERT}  \cite{devlin2018bert} were designed to be pre-trained on massive corpora of text, but fine-tuned for some specific task. For example, \textit{SciBERT} \cite{beltagy2019scibert} is a BERT-based model that is fine-tuned on the full text of a large sample of research papers from Semantic Scholar. 

In the field of Network Analysis, \textit{node embedding} refers to the task of learning low-dimensional vector representations for the set of nodes on a graph \citep{cai2018comprehensive,goyal2018graph}. These methods are primarily designed to preserve \emph{proximity}, such that similar embeddings are learned for nodes which might belong to the same region or community in the graph. 
The \emph{DeepWalk} \cite{perozzi2014deepwalk} method, which extends the NLP SkipGram model \citep{mikolov2013efficient}, learns to predict those nodes that will co-occur on random walks on the graph,  so that the resulting representations preserve proximities between nodes.

\textit{SPECTER} \cite{cohan2020specter} uses citation relations between papers as an external \textit{inter-document signal} to fine tune text-based representations. Initialised with SciBERT and using title and abstract text from articles, SPECTER is trained to minimise the distance between the representations of a query paper and an article that it cites, while maximising the distance between the query paper and an article that it does not cite. 

Message Passing Graph Neural Networks (MP-GNNs) (e.g. \textit{GraphSAGE} \cite{hamilton2017inductive}, \textit{GCN} \cite{kipf2016semi}, \textit{VGAE} \cite{kipf2016variational}) can be applied to graph structured data, where some additional information is available for the nodes. These methods can then learn node embeddings which depend not only on a node's descriptive features, but also the features of its neighbours. MP-GNNs iteratively update the representation of each node by receiving information from neighbouring nodes and learning aggregation functions to combine those messages with a node's current representation. In the context of a citation network, where a node's features are typically derived from the text of the corresponding paper, the representation of some focal paper is then informed not only by its own content, but that of the papers that it cites (the ideas upon which it develops), and even the articles that cite the focal paper (its applications or the work that builds on it). 

In the context of interdisciplinary research papers, these may present in citation graphs as \textit{bridges} or \textit{bottlenecks} with both homophilic and heterophilic connections. This makes the task of representing interdisciplinary research more difficult from a Graph Learning perspective. Modelling bottleneck structures and heterophilic interactions remain open challenges in the field, hampered by representation learning phenomena like \textit{over-squashing} \cite{topping2021understanding} and \textit{over-smoothing} \cite{yan2021two}. Therefore, in this paper we further develop the concept of MP-GNNs and propose a novel method of message aggregation, such that important local graph structures (e.g. co-citations) can be considered in an effort to learn article embeddings that will provide better representations for interdisciplinary works.

\section{Methods}

\begin{figure*}[h]
  \includegraphics[width=0.75\textwidth]{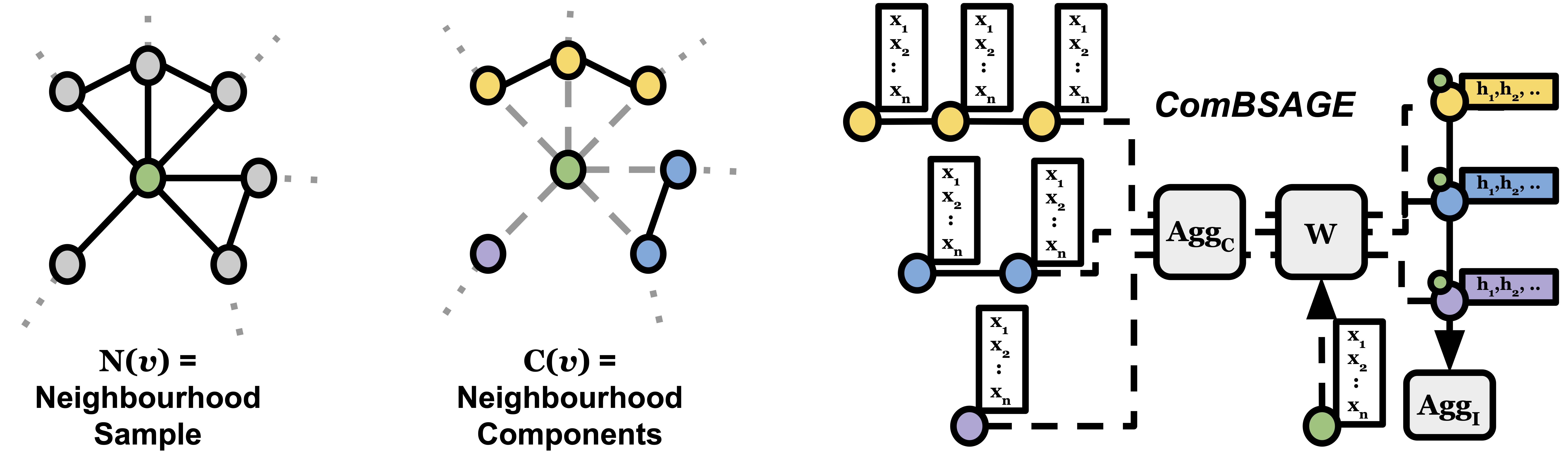}
  \caption{ComBSAGE. Messages are grouped accord to $C$, and aggregated separately by $AGG_C$, before being combined by $AGG_I$.}
    \vskip-1em
  \label{fig:method}
\end{figure*}

In this section we present our novel Community-Based Sample and Aggregation (ComBSAGE) method for Message Passing Graph Neural Networks. The goal of this architecture is to incorporate local structural information in the message aggregation stage of MP-GNNs, in order to provide scientific document embeddings that are more appropriate when representing interdisciplinary research, without compromising the quality of paper representations overall.

The motivation behind this approach is as follows. Suppose we wish to aggregate messages from three neighbouring nodes, two of which are connected. We propose that these messages should be combined in a manner that accounts for this connection. Consider, for example, the task of representing a research paper which draws heavily on methods from the fields of mathematics and computer science, and that has a few recent applications in a field like political science. Such a paper presents in a citation network as a node with many connections to a large community of mathematics and computer science-related publications, and a smaller number of connections to works in political science. A traditional MP-GNN may lose much of the signal from the political science papers when aggregating messages. This relates to the \textit{over-smoothing} phenomenon that has been highlighted as a weakness with many existing GNN architectures; where node representations converge and become indistinguishable from their neighbours \cite{chen2020measuring}, and equally to the \textit{over-squashing} problem; where information is lost in nodes that represent bottlenecks in the network \cite{topping2021understanding}. Our proposed method of message aggregation splits messages from distinct communities, such that they can be aggregated separately before being combined. We suggest that this approach allows the important interdisciplinary implications of a research paper (e.g. its applications in political science) to be preserved in its representation.

One layer of an MP-GNN is typically described in two functions: \textit{aggregate} and \textit{update}. For a node $v_i$, with features described by a vector $x_i$, we denote the neighbours of $v_i$ as $\{v_j,v_k, ... ,v_p\}$ with features $\{x_j,x_k, ... ,x_p\}$. An embedding for $v_i$ (denoted by $h_i$) is then generated via the update $h_i \gets \phi(x_i, u_i)$, where $u_i \gets \psi(\{x_j,x_k, ... ,x_p\}, x_i)$. Thus, $\psi$ is some (typically permutation invariant) function for \textit{aggregating} information from neighbouring nodes (potentially with consideration of $x_i$), and $\phi$ is some parametric function for \textit{updating} the representation of a node using the aggregated message. MP-GNNs can be trained in a supervised fashion (e.g. in pursuit of some task such as node classification), or in an unsupervised manner where the objective task used for training is graph reconstruction or link prediction. 

\begin{algorithm}[!t]
\SetKwInOut{Input}{Input}
\SetKwInOut{Output}{Output}
\caption{Community-based Aggregation}
\label{alg:combsage}
    \Input{Graph $G(V,E)$; input features $\{ x_v, \forall v \in V \}$; \newline weight matrices $W_I$, $W_C$;
    non-linearity $\sigma$; \newline differentiable aggregator functions $AGG_C^k$ and $AGG_I^k, \forall k \in 1...K$; neighbourhood function $N: v \rightarrow 2^v$; neighbourhood component function $C: v \rightarrow \{c, c \subseteq N(v)\}$}
    \Output{Vector representations $z_v$ for all $v \in V$}
    \For{$k = 1...K$}{
    \For{$v \in V$}{
    \For{$c \in C(v)$}{
    $h_c^k \gets AGG_C^k(\{h_u^{k-1}, \forall u \in C\})$\;
    $h_c^k \gets \sigma( W_I \cdot CONCAT(h_v^{k-1}, h_c^k))$\;
    }
    $h_v^k \gets AGG_I^k(\{ h_c^k, \forall c \in C\})$\;
    $h_v^k \gets \sigma( W_C \cdot CONCAT(h_v^{k-1}, h_v^k))$\;
    }
    }
    $z_v \gets h_v^K, \forall v \in V$
\end{algorithm}

The proposed ComBSAGE method is outlined in Algorithm \ref{alg:combsage}. The key contribution of our approach is the aggregation function ($\psi$), in lines 3 -- 7 of Algorithm \ref{alg:combsage}.
The neighbourhood component function $C$ partitions the set of neighbours of some node $v$ into sets of connected components, i.e. $C(v)$ represents the connected components of the subgraph of $G$ induced by the set $N(v)$, (the neighbours of $v$). As such, if some node $v$ is part of a fully connected clique, then $C(v)$ contains only one set -- $N(v)$. Conversely, if $v$ represents a hub node connecting $n$ otherwise isolated neighbours, then $C(v)$ contains $n$ sets, each containing a single neighbour. The ComBSAGE method then applies 
message aggregations first within each of the components of $C(v)$, and then between them. We illustrate the operation of the neighbourhood component function and a single layer of the ComBSAGE aggregation process in Figure \ref{fig:method}.  

\section{Data}
\label{sec:data}
To evaluate our proposed method, we require a dense, connected citation network that contains research from a diverse set of scientific disciplines. We construct such a graph using Semantic Scholar citation information for 58,513 research papers. In order to ensure that the graph contains regions of interdisciplinary research, we collect the research articles according to the following process:
(i) Select 8 topics from the All Science Journal Categorisation (ASJC): \emph{`Computer Science'}, \emph{`Mathematics'}, \emph{`Chemistry'}, \emph{`Medicine'}, \emph{`Social Sciences'}, \emph{`Neuroscience'}, \emph{`Engineering'}, and \emph{`Biochemistry, Genetics and Molecular Biology'}.
(ii) From each ASJC, select a random sample of up to 1,500 articles published in journals assigned to that topic. 
(iii) Collect any additional articles that have a `cited by' or `citing' relationship with at least 10 articles in the seed set.  
(iv) Filter the resulting network: remove any articles for which abstract text is not available in English, and retain only the largest connected component in the network. 
In this manner, we collect a dense, multidisciplinary citation graph with title and abstract text for each article. In the case of each paper, we encode the concatenated title and abstract using the pre-trained SciBERT model and use the final representation of the [CLS] token as an embedding to represent each document (similar to the approach in \cite{cohan2020specter}). The resulting embedding is stored as the only metadata considered on each node. 

\section{Evaluation}

We implement six methods of paper representation for evaluation: (i) \textit{DeepWalk} which does not consider paper content; (ii) \textit{SciBERT} where no citations relations are known between papers; (iii) \textit{SPECTER}; (iv) \textit{GraphSAGE} (MEAN) with a non-parametric mean aggregator; (v) \textit{GraphSAGE} (LSTM) with a learned LSTM aggregator; (vi) our proposed method \textit{ComBSAGE}.
Methods (i) and (iii) are trained on our data using the default hyperparameters. (ii) is a pre-trained model from AllenAI. For the GNN methods (iv -- vi), each model is implemented with 2 layers, and the batch size, node sampling, and learning rate hyperparameters are all fixed using the best performing set identified in an exhaustive grid-search with the GraphSAGE (MEAN) model. For the ComBSAGE model, we choose a max-pooling aggregator for $AGG_C$ and an LSTM aggregator for $AGG_I$. We also include a \textit{jumping knowledge} \cite{xu2018representation} connection between the hidden representations in the first and second layer, as this has been shown to be effective in low-homophily networks. 

We evaluate each representation method (or paper embedding) based on a citation prediction task. From the data described in Section \ref{sec:data} above, we randomly sample 20\% and 5\% of the citations to serve as test and validation sets respectively. We learn embeddings from the remaining citation graph according to each of the chosen methods. To treat citation prediction as a binary classification task, we supplement each set of citations (train, validation, and test) with negatively sampled edges/citations that did not occur in the graph. Each edge is then represented as the Hadamard or element-wise product between the embeddings of its endpoints, and classified using a Multi-Layer Perceptron (MLP). We repeat this process for 5 random train:validation:test splits. The overall performance of each embedding approach is reported in Table \ref{tab:overall} in Section \ref{sec:results}.

\begin{SCfigure}
    \includegraphics[width=0.23\textwidth]{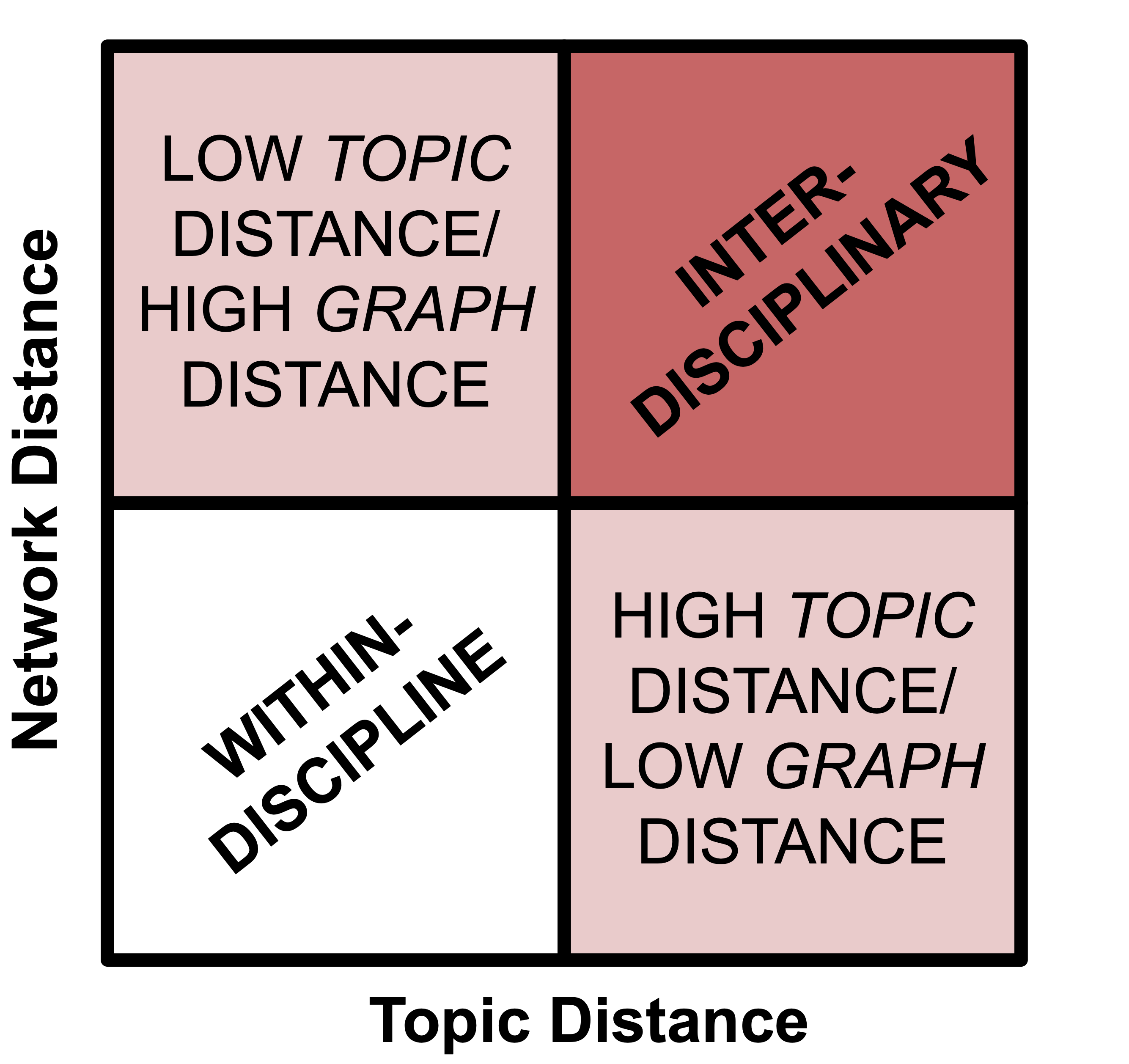}
    \caption{Quadrant of citation distance. We consider a citation to be \textit{interdisciplinary} if it is between two papers that are distant according to their topic distance (relating to the dissimilarity of their content), and their network distance (relating to their positions in the citation network).}
    \label{fig:quad}
\end{SCfigure}


In addition to reporting the performance across the entire citation graph, we also wish to understand how each method performs in the interdisciplinary regions of the citation network. Accordingly, we adopt two definitions of \textit{distance} for citations: (i) the \textit{network} distance of a citation, which is the distance between two papers in the citation graph prior to the citation; (ii) the \textit{topic} distance of a citation, which is the dissimilarity between the content of the papers. Using these definitions of distance, every citation can be considered on the quadrant described in Figure \ref{fig:quad}, with the most interdisciplinary citations positioned in the upper right corner, corresponding to citations between papers positioned in distant regions (i.e., with minimal overlap between their references and citations {\textbf{and}} with minimal similarity between the content of the research papers). In practice, we measure network and topic distance of citations as the cosine distance between the DeepWalk and SciBERT embeddings of the endpoints respectively. We divide all citations in the test set into the four regions of this quadrant and report how paper embedding methods perform at predicting citations in each region. These results are presented in Figure \ref{fig:idr}.

\section{Results}
\label{sec:results}

Table \ref{tab:overall} reports the overall citation prediction performance of each method as measured using the Area Under the ROC Curve. We find that methods which leverage both document text and citation graph information outperform methods which rely on only one modality. Further, we note that our proposed method (\textit{ComBSAGE}) is competitive with the state of the art, with GNN-based methods outperforming more NLP-focused approaches (e.g. \textit{SPECTER}).

\begin{table}[h]
    \centering
    \vskip-1em
    \begin{tabular}{llr}
        \toprule
        {} & Method &            Overall AUC \\
        \midrule
      (i) &\textit{DeepWalk} \cite{perozzi2014deepwalk}       &  0.95+0.0004 \\
      (ii) &\textit{SciBERT} \cite{beltagy2019scibert}        &  0.93+0.0002 \\
      (iii) &\textit{SPECTER} \cite{cohan2020specter}        &  0.95+0.0048 \\
        (iv) &\textit{GraphSAGE (LSTM)} \cite{hamilton2017inductive}  &  0.97+0.0037 \\
        (iv)& \textit{GraphSAGE (MEAN)} \cite{hamilton2017inductive}      &  \textbf{0.98+0.0003} \\
        (vi) &\textit{ComBSAGE} (proposed) &  \textbf{0.98+0.0016} \\
        \bottomrule
    \end{tabular}
    \caption{Citation prediction performance in terms of Area Under the ROC Curve (AUC). The best performing methods are marked in bold.}
    \vskip-2em
    \label{tab:overall}
\end{table}

\begin{figure}[!t]
  \centering
  \includegraphics[width=0.38\textwidth]{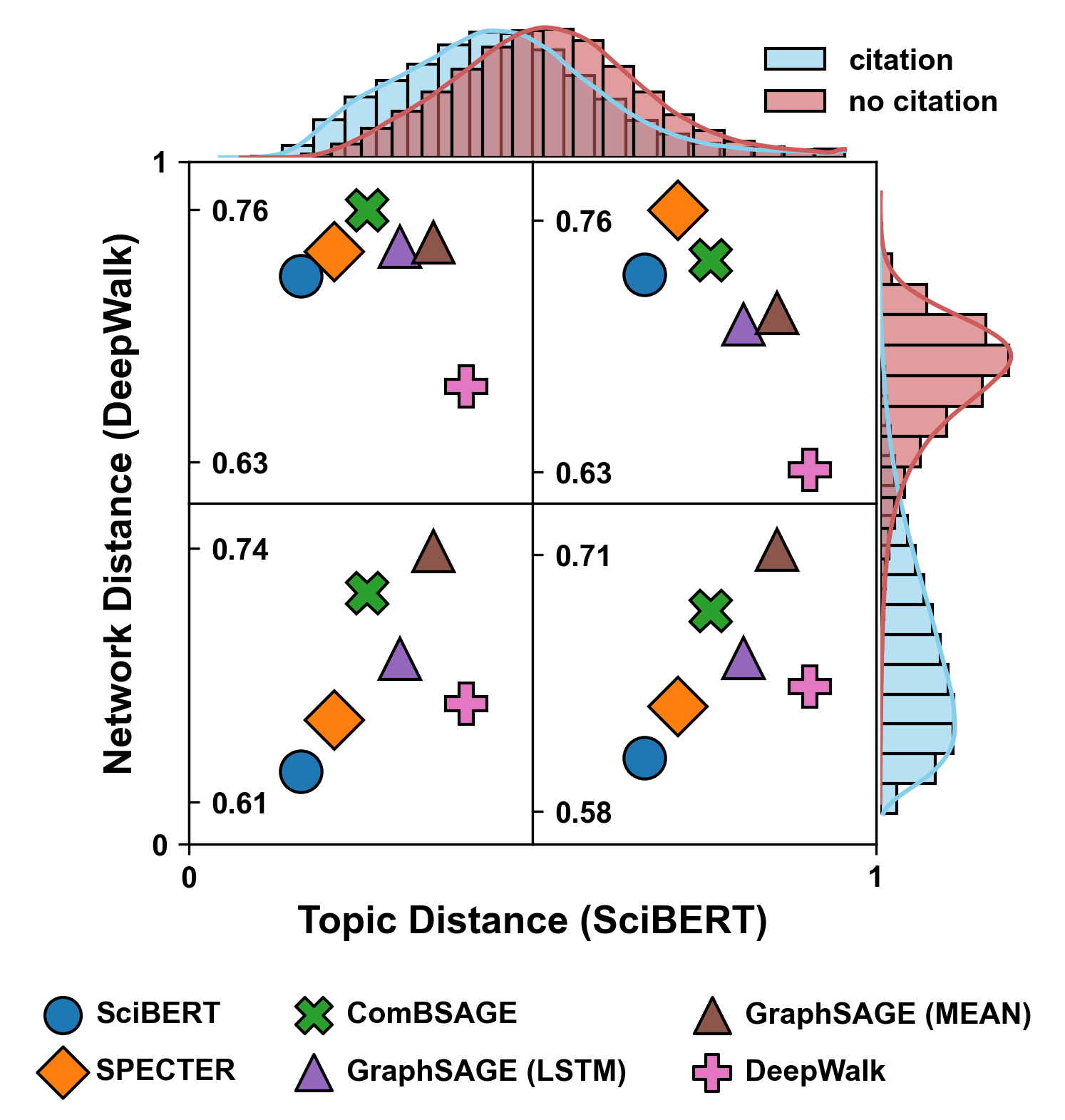}
\vskip -1.2em
  \caption{Citation prediction performance (mean class accuracy), with citations grouped according to the distance between the papers. }
  \label{fig:idr}
\vskip -2em
\end{figure}

Figure \ref{fig:idr} shows the balanced accuracy score for citation prediction for each embedding method, in each region of the citation distance quadrant (as per Figure \ref{fig:quad}). The $y$ position of a marker within each subsection shows the average balanced accuracy score over 5 splits of the data. The markers have been jittered along the $x$-axis for legibility. While the order of the markers along this axis could be considered arbitrary, we have attempted to position methods according to their position on the spectrum in Figure \ref{fig:spectrum}. It is apparent from the bottom half of this diagram that citation information is highly important in understanding relations between papers with overlapping citation neighbourhoods. Conversely, when identifying relations between papers that are distant in the citation network, the utility of textual features increases. Our proposed ComBSAGE method outperforms other GNN-based methods when predicting interdisciplinary citations (see also Appendix). 

\section{Discussion \& Conclusions}

In this work, we evaluate a variety of document embedding methods on their ability to identify interdisciplinary relations between research papers. We find that those methods that rely more on NLP and the textual content of research papers (\textit{SciBERT}, \textit{SPECTER}) are better able to predict interdisciplinary citations than methods which rely more on network analysis of citation information (\textit{GraphSAGE}, \textit{DeepWalk}). However, these NLP-based methods are not sufficiently capable of leveraging citation information to compete with GNN-based methods in predicting interactions between papers with overlapping citation neighbourhoods or in the task of citation prediction \textit{overall}. Motivated by recent work in citation analysis of interdisciplinary research, we propose a novel method of GNN message aggregation, \textit{ComBSAGE}, that rectifies some of the limitations of MP-GNNs when predicting \textit{interdisciplinary} citations, without compromising \textit{overall} citation prediction performance. 

\vskip 0.2em
\noindent\textbf{Acknowledgements.} This research was supported by Science Foundation Ireland (SFI) under Grant Number SFI/12/RC/2289\_P2. 

\bibliographystyle{ACM-Reference-Format}
\bibliography{references}
\clearpage
\onecolumn
\appendix

\section*{Appendix}
\begin{table*}[htbp!]
    \centering
\begin{tabular}{llll}
\toprule
Method & AUPRC & F1 (macro average) & Average Precision \\
\midrule
(i) \textit{DeepWalk} & 0.345±0.003 & 0.640±0.001 & 0.198±0.002 \\
(ii) \textit{SciBERT} & 0.379±0.004 & 0.654±0.001 & 0.239±0.002 \\
(iii) \textit{SPECTER} & 0.515±0.058 & 0.710±0.020 & 0.303±0.027 \\
(iv) \textit{GraphSAGE (LSTM)} & 0.441±0.026 & 0.698±0.010 & 0.273±0.013 \\
(v) \textit{GraphSAGE (MEAN)} & 0.531±0.003 & 0.730±0.002 & 0.324±0.003 \\
(vi) \textit{ComBSAGE} (proposed) & \textbf{0.543±0.008} & \textbf{0.738±0.003} & \textbf{0.335±0.006} \\
\bottomrule
\end{tabular}
    \caption{Interdisciplinary citation prediction performance according to different metrics. We report Area under the Precision Recall Curve (AUPRC), macro averaged F1 score and Average Precision (AP). In particular, we choose AUPRC and AP as metrics which are especially informative/reliable when the positive class represents a small minority in the dataset; as is the case in the IDR region of our test set.}
    \vskip-2em
    \label{tab:idr}
\end{table*}

\end{document}